\documentclass[12pt]{article}
\usepackage{amssymb,amsthm}
\usepackage{amstext}
\usepackage{amsmath}
\usepackage{array}
\usepackage{amscd}
\usepackage{xypic}
\usepackage{bm}

\newtheorem{exam}{Example}

\newcommand{\nn}{\nonumber}
\newcommand{\AB}{\allowbreak}
\newcommand{\gr}[2]{\mathop{{\mathbf{#1}}(#2)}\nolimits}

\newcommand{\ali}[2]{\mathop{\mathfrak{#1}(#2)}\nolimits}

\newcommand{\ad}{\mathop{\mathrm{ad}}\nolimits}
\newcommand{\ADA}[1]{\ifmmode \ad(#1) \else $\ad(#1)$\fi}
\newcommand{\LI}[2]{\ifmmode#2_1,\AB\,\ldots,\,\AB #2_{#1}%
\else$ #2_1,\AB\,\ldots,\,\AB#2_{#1}$\fi}

\newcommand{\su}[1]{\ali{su}{#1}}
\newcommand{\sltwo}{\ifmmode \ali{sl}{2} \else $\ali{sl}{2}$\fi}

\long\def\comment#1{}

\newcommand{\bMA}[1]{\[\begin{array}{#1}}
\newcommand{\eMA}{\end{array}\]}

\newcommand{\C}{{{\mathbb C}}}

\newcommand{\NR}{{\mathbb{R}}}

\def\be{\begin{equation}}
\def\ee{\end{equation}}

\def\R{\NR}

\def\cpn{{\C P^{N-1}}}
\def\bp{{\bar{\partial}}}
\def\p{{\partial}}

\def\tr{{\mathrm{tr}}}

\mathchardef\za="710B  
\mathchardef\zb="710C  
\mathchardef\zg="710D  
\mathchardef\zd="710E  
\mathchardef\ze="710F  
\mathchardef\zz="7110  
\mathchardef\zh="7111  
\mathchardef\zy="7112  
\mathchardef\zi="7113  
\mathchardef\zk="7114  
\mathchardef\zl="7115  
\mathchardef\zm="7116  
\mathchardef\zn="7117  
\mathchardef\zx="7118  
\mathchardef\zp="7119  
\mathchardef\zr="711A  
\mathchardef\zs="711B  
\mathchardef\zt="711C  
\mathchardef\zu="711D  
\mathchardef\zf="711E  
\mathchardef\zq="711F  
\mathchardef\zc="7120  
\mathchardef\zw="7121  
\mathchardef\zG="7000  
\mathchardef\zD="7001  
\mathchardef\zY="7002  
\mathchardef\zL="7003  
\mathchardef\zX="7004  
\mathchardef\zP="7005  
\mathchardef\zS="7006  
\mathchardef\zU="7007  
\mathchardef\zF="7008  
\mathchardef\zC="7009  
\mathchardef\zW="700A  
\mathchardef\ze="7122  
\mathchardef\zvy="7123  
\mathchardef\zvr="7125 
\mathchardef\zvs="7126 
\mathchardef\zvf="7127  

\newcommand{\bea}{\begin{equationarray}}
\newcommand{\eea}{\end{equationarray}}
\newcommand{\la}{\lambda}
\renewcommand{\c}{\cdot}

\begin{document}





\title{Invariant recurrence relations for $\cpn$ models
}

\author{}


\maketitle
\begin{flushleft}
{\bf P. Goldstein$^1$ and A. M. Grundland$^2$}\\
$^1$ Theoretical Physics Department, \\
The Andrzej Soltan Institute for Nuclear Studies, \\
Hoza 69, 00-681 Warsaw, Poland\\
Piotr.Goldstein@fuw.edu.pl\\
\vspace{5mm}
$^2$ Centre de Recherches Math{\'e}matiques, Universit{\'e} de Montr{\'e}al, \\
C. P. 6128, Succ.\ Centre-ville, Montr{\'e}al, (QC) H3C 3J7, Canada \\
Universit\'{e} du Qu\'{e}bec, Trois-Rivi\`{e}res CP500 (QC) G9A 5H7, Canada\\
grundlan@crm.umontreal.ca
\end{flushleft}

\begin{abstract}
In this paper, we present invariant recurrence relations for the
completely integrable $\cpn$ Euclidean sigma model in two
dimensions defined on the Riemann sphere $S^2$ when its action
functional is finite. We determine the links between successive
projection operators, wave functions of the linear spectral
problem, and immersion functions of surfaces in the $\su{N}$
algebra together with outlines of the proofs. Our formulation
preserves the conformal and scaling invariance of these
quantities. Certain geometrical aspects of these relations are
described. We also discuss the singularities of meromorphic
solutions of the $\cpn$ model and show that they do not affect the
invariant quantities. We illustrate the construction procedure
through the examples of the $\mathbb{C}P^2$ and $\mathbb{C}P^3$
models.
\end{abstract}

\section{Introduction}
\label{sec:Intro}

The general properties of the $\cpn$ models and techniques for
finding associated surfaces remain among the essential subjects of
investigation in modern mathematics and physics. In the case of the
completely integrable Euclidean $\cpn$ sigma model in two dimensions
an efficient and still useful approach has been the use of the Lax
pair as introduced by Zakharov and Mikhailov \cite{Mik,MikZ}.
Especially fruitful was geometrization of the spectral theory by
representing the equations as conditions for the immersion of
surfaces in multidimensional Euclidean spaces. Based on the linear
spectral problem for integrable equations, the concept of
constructing infinitely many surfaces immersed in multidimensional
spaces was first presented by Sym and Tafel (ST)
\cite{Sym1,Sym2,Sym3,Taf}. The advantage of their formula is that it
allows us to express an immersion function of a surface directly in
terms of a wave function satisfying the associated linear spectral
problem. This subject has been developed further by many authors (see
e.g. \cite{Bob,Fok,FokG,HelW,KL,Kono2,RSW,Uhl,GSZ} and references
therein). More recently, conservation laws of the considered model
have led to the generalized Weierstrass formula for immersion (GWFI)
of 2D surfaces, which originated from the work of Konopelchenko
\cite{Kono1}. It was shown \cite{GY1,GY2} that for 2D surfaces
immersed in $\su{N}$ algebras in the case of $\cpn$ models the ST
formula coincides with the GWFI. On the basis of this geometrical
view the present authors focus their attention on the case where the
model is defined on the Riemann sphere $S^2$ and its action
functional is finite. The complete set of regular solutions is known
due to Din and Zakrzewski as well as Sasaki, Eells and Wood
\cite{Din,Sa,EW}. Under the above assumptions, the considered
surfaces are conformally parametrized. In the classical approach to
the $\cpn$ models \cite{Din} new solutions are constructed by
multiple application of a ``creation operator'' $P_+$ to any
holomorphic solution or an ``annihilation operator'' $P_-$ to any
antiholomorphic solution.

 It seems worthwhile to provide an invariant
formulation of the main ingredients of the theory. Namely, the
considered models are complex projective. The equations of motion as
well as their integration schemes are invariant under scaling not
only by a constant factor but by any scalar complex-valued function.
For such models the natural approach seems to be expressing all the
quantities in scaling-invariant form. In this paper we formulate the
theory in terms of invariant projection operators rather than the
previously used unnormalized (homogeneous) coordinates.

Starting from the invariant Lagrangian density we regain the
well-known equations of motion in the form of conservation laws
\cite{Mik,MikZ} and then we construct the solutions in a way similar
to \cite{Din} by means of the appropriate ``creation'' and
``annihilation'' operators applied to those projectors. The
corresponding operators are also derived for the wave functions of
the spectral problem and for the immersion functions of surfaces
corresponding to those functions (soliton surfaces). Finally, the
geometrical characteristics of the surfaces are also expressed in
terms of the projectors. We complete the approach by comments on
possible behavior of the invariant solutions in a neighborhood of
(what used to be) singularities of homogeneous field coordinates in
the $\cpn$ model. For a deeper insight into the $\cpn$ model
theories, we refer the reader to some standard books in the field
\cite{Bob2,Bab,GU2,HelF,Ken2,MS,Osser,WZ}.

Throughout this paper we use the terms ``creation'' and
``annihilation'' operators, suggested by the commonly applied symbols
$P_\pm$. However the reader should bear in mind that the procedure is
a walk over a sphere rather than up or down a ladder (for this reason
we retain the quotation marks). The construction of orthogonal
functions or projectors is in fact an application of the classical
Gram-Schmidt orthogonalization procedure \cite{LL} in which the
subsequent base vectors are constructed from derivatives of its
predecessors. This aspect will later be discussed in more detail.

 This paper is organized as follows. In section 2 we recall the
main elements of the $\cpn$ theory, which will be the basis for
further calculations (this includes the introduction of the invariant
Lagrangian). Section 3 is devoted to description of the invariant
recurrence formulae for the $\cpn$ models. The goals are described in
diagram 1.

Diagram 1 Relations between projectors, wave functions and
immersion functions associated with the $\cpn$ model

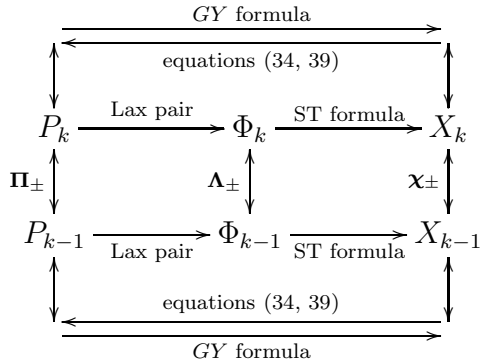
$$
\xymatrix@C=44pt{
& \ar@{<->}[d] \ar@<0.5ex>[rr]^{\text{$GY$ formula}} &
&\ar@<0.5ex>[ll]^{\text{equations (\ref{Pk from Xk}, \ref{PkX})}} \ar@{<->}[d] \\
& P_k \ar@{->}[r]^{\text{Lax pair}} \ar@{<->}[d]_{{\bm \Pi}_{\pm}}
& \Phi_k \ar@{->}[r]^{\text{ST formula}} \ar@{<->}[d]_{\bm
{\Lambda}_{\pm}}
& X_k \ar@{<->}[d]_{{\bm \chi}_{\pm}}\\
& P_{k-1} \ar@{->}[r]_{\text{Lax pair}} & \Phi_{k-1} \ar@{->}[r]_{\text{ST formula}} & X_{k-1}\\
& \ar@{<->}[u] \ar@<-0.5ex>[rr]_{\text{$GY$ formula}} & &
\ar@<-0.5ex>[ll]_{\text{equations (\ref{Pk from Xk}, \ref{PkX})}}
\ar@{<->}[u] }
$$

\noindent We seek the link between the quantities $P_k$ and
$P_{k-1}$, $\Phi_k$ and $\Phi_{k-1}$, $X_k$ and $X_{k-1}$. It
should be noted that from the diagram above, the projectors $P_k$
and $P_{k-1}$ are related to the wave functions $\Phi_k$ and
$\Phi_{k-1}$ respectively through the concept of Lax pairs.
Likewise, the projectors $P_k$ and $P_{k-1}$ are connected with
the immersion functions $X_k$ and $X_{k-1}$ respectively through
the GY formula provided by the second author of this paper
\cite{GY1}. In our formulation we preserve the conformal
invariance of the action functional, projectors, wave functions,
surfaces under consideration. The derivation of the ``creation''
and ``annihilation'' operators for projectors $P_k$, wave
functions $\Phi_k$ and surfaces $X_k$ are put off until the
appendices. The above procedure is illustrated by means of several
examples (namely the $\mathbb{C}P^2$ and $\mathbb{C}P^3$ models).
In section 4 we discuss some geometrical characteristics for the
recurrence relations between consecutive surfaces in terms of the
previous ones. Section 5 comments on the singularity structure of
solutions of the $\cpn$ model. We show that singularities of
meromorphic solutions of the model do not extend to invariant
quantities. Section 6 summarizes the obtained results and contains
some suggestions regarding further developments.

\section{Basic facts and notions on the $\cpn$ model}
To make the paper self-contained we briefly summarize the basic facts
on the $\cpn$ model theory which constitute the background of our
calculations.

Dynamics of the $\cpn$ sigma models defined on the Riemann sphere
$S^2$ is determined by stationary points of the action functional
(see e.g. \cite{WZ})
\be
\label{action} S=\int\!\!\!\int_{S^2}\mathcal{L} d\xi
d\bar{\xi}=\frac{1}{4}\int\!\!\!\int_{S^2}(D_{\mu}z)^{\dagger}\cdot(D_{\mu}z)d\xi
d\bar{\xi},
\ee
where the Lagrangian density $\mathcal{L}$ is
\be
\mathcal{L} = \frac{1}{4}(D_{\mu}z)^{\dagger}\cdot(D_{\mu}z),
\label{lagr-z}
\ee
and the covariant derivatives $D_{\mu}$ are defined according to the formula
\be
\label{cov}
D_{\mu}z=\partial_{\mu}z-(z^{\dagger}\cdot\partial_{\mu}z)z,\qquad\partial_{\mu}
=\partial_{\xi^{\mu}},\qquad\mu=1,2.
\ee
The field variables $z=(z_0,...,z_{N-1})$ are
points of the coordinate space which is the $(N-1)$-dimensional unit
sphere immersed in $\C^N$
\be
\label{normalization}
z^\dagger\c z=1,
\ee
with the usual definition of the scalar product, while $z^\dagger$
is the Hermitian conjugate of $z$. The space of independent
variables is two dimensional. Originally being also the unit
sphere, it is usually converted to the Riemann sphere $S^2=\C\cup\
\{\infty\}$ by stereographic projection. In our paper the
independent variables are pairs $(\xi^1,\xi^2)~\in \R^2$ or
$(\xi,\bar{\xi})~\in \C$, with $\xi=\xi^1+i\,\xi^2$, where complex
conjugates are marked by a bar over a quantity.

The normalization of $z$ \eqref{normalization} imposes a constraint
on its components, which makes them inconvenient. The common approach
to the $\cpn$ models is to describe the models in terms of the
homogeneous, unnormalized field variables $f$, such that
$z=f/(f^\dagger\c f)^{1/2}$. The vector $z$ is determined by the
Euler-Lagrange (E-L) equations with the constraints
(\ref{normalization})
\be
D_{\mu}D_{\mu}z+(D_{\mu}z)^{\dagger}\cdot(D_{\mu}z)z=0,\qquad z^{\dagger}\cdot z=1,
\ee
whereas the homogeneous variables $f$ satisfy an unconstrained
form of the E-L equations
\be
\label{E-L} \left(\mathbb{I}-\frac{f\otimes
f^{\dagger}}{f^{\dagger}\cdot f}\right)\cdot\left[\p\bp
f-\frac{1}{f^{\dagger}\cdot f}\left((f^{\dagger}\cdot\bp f)\p
f+(f^{\dagger}\cdot\p f)\bp f\right)\right]=0,
\ee
where $\p$ and $\bp$ denote the derivatives with respect to $\xi$
and $\bar{\xi}$ respectively and $\mathbb{I}$ is the $N\times N$
unit matrix. An important property of these equations is their
invariance under scaling by multiplication of $f$ by an arbitrary
scalar function $\varphi(\xi)$.

The E-L equations (\ref{E-L}) take the elegant form of a
conservation law if we express them in terms of Hermitian
projection matrices $P:S^2\rightarrow \mbox{Aut}(\mathbb{C}^N)$
\be
P=(1/f^\dagger\c f)f\otimes f^{\dagger},\qquad P^2=P,\qquad P^{\dagger}=P,
\ee
namely
\be\label{cons-law}
\p\,[\bp P,P]+\bp\,[\p P,P]=0
\ee
In this paper we are going to use the projectors $P$ as our
fundamental unknown variables. The advantage of such an approach is
the explicit invariance of these variables under scaling with any
scalar function of $\xi$. Thus the scaling-invariant Euler-Lagrange
equations \eqref{cons-law} are expressed in scaling-invariant
variables. On the other hand, the projectors are obviously subject to
another constraint: $P^2=P$. Due this constraint we introduce the
Lagrange multiplier
$\lambda=\lambda^{\dagger}\in\mbox{Aut}(\mathbb{C}^N)$ into the
action (\ref{action}) and we get
\be
\label{action2}
S=\int_{S^2}\tr{\{\p P\cdot\bp P+\lambda\cdot(P^2-P)\}}d\xi d\bar{\xi}.
\ee
By the variation of the action (\ref{action2}) we obtain
\be
\label{variation1}
\begin{split}
\delta\lambda&:\quad P^2-P=0\\
P&:\quad 2\p\bp P+\lambda\cdot P+P\cdot\lambda-\lambda=0
\end{split}
\ee
We eliminate the Lagrange multiplier $\lambda$ by multiplying
(\ref{variation1}) from the left and from the right by $P$. Next
we subtract the obtained results which yields the equation
(\ref{cons-law}) as the E-L equation of the action
(\ref{action2}).

The conservation law \eqref{cons-law} means that the 1-form
\be\label{dX}
dX = i\left(-[\p P,P]d\xi+[\bp P,P]d\bar{\xi}\right)
\ee
is a closed differential. Hence its integral, independent of a
trajectory, may be used to construct the following $N\times N$ matrix
in $\su{N}$
\be\label{X}
X(\xi,\bar{\xi})=i\int_{\gamma}\left(-[\p P,P]d\xi+[\bp P,P]d\bar{\xi}\right),
\ee
which may be regarded as a surface immersed in a real
$(N^2-1)$-dimensional space \cite{GSZ}. The mapping
$X:S^2\ni(\xi,\bar{\xi})\rightarrow X(\xi,\bar{\xi})\in\su{N}$ is
known in the literature \cite{Kono1,NS} as the generalized
Weierstrass formula for immersion of 2D surfaces in
$\mathbb{R}^{(N^2-1)}\cong\su{N}$. The space is equipped with the
scalar product
\be\label{scalar}
(A,B)=-(1/2) \tr(A\c B),\quad A,~B\in \su(N)
\ee
which is used to construct an orthonormal basis
(Pauli matrices in 3 dimensions, Gelfand matrices in 8 dimensions,
etc. see e.g. \cite{Hol,Til}).

In a classical paper \cite{Din} a base of vectors $f_i$ was
constructed by the Gram-Schmidt orthogonalization, namely consecutive
applications of the contracting operator $P_+$ defined by
\be
\label{P+}
P_+(f)=(\mathbb{I}-P)\c \p f,
\ee
which we will refer to as a ``creation operator'',
while the inverse operation $P_-$, an ``annihilation operator'', is defined by
\be\label{P-}
P_-(f)=(\mathbb{I}-P)\c\bp f.
\ee
The usual procedure of constructing the orthogonal basis
$f_0,...,f_{N-1}$ includes normalization by setting the first nonzero
component of each $f_k$ to one. Unlike the standard creation and
annihilation operators known in the literature \cite{BD,LL}, the
operations (\ref{P+}, \ref{P-}) leading to the new vectors $f$ are
nonlinear. Although the new vectors are obtained from their
predecessors by a linear operation of matrix multiplication, the
multiplier, which defines the direction of the projection, also
depends on the argument (the new direction is the projection of the
tangent to the graph $f(\xi,\bar{\xi})$ onto the hyperplane
orthogonal to the vector $f(\xi,\bar{\xi})$).

It was shown in \cite{Din} that multiple applications of $P_+$ to any
holomorphic function lead to an antiholomorphic one after $(N-1)$
steps and obviously the application of $P_+$ to an antiholomorphic
function yields zero. This way we obtain $N$ orthogonal functions
${f_0,...,f_{N-1}}$, and -- as a by-product -- $N$ projectors
$P_0,...,P_{N-1}$ acting on orthogonal complements of one-dimensional
subspaces in $\mathbb{C}^N$.

In \cite{MikZ} the linear problem containing a spectral parameter
$\la\in\mathbb{C}$ was found in the form of a system
\be\label{spectral}
\p \Phi_k=\frac{2}{1+\la}[\p P_k,P_k]\Phi_k , \qquad \bp
\Phi_k=\frac{2}{1-\la}[\bp P_k,P_k]\Phi_k, \qquad
k=0,1,\ldots,N-1,
\ee
where $\la\in\mathbb{C}$ is the spectral parameter and $P_k$ is a
sequence of rank-1 orthogonal projectors which map on the direction
of $f$
\be
\label{projk} P_k=\frac{f_k\otimes f_k^{\dagger}}{f_k^{\dagger}\cdot
f_k},\qquad f_k=P_{\pm}^kf,\qquad P_k^2=P_k,\qquad P_k^{\dagger}=P_k.
\ee
The compatibility condition for equation (\ref{spectral}) 
correponds precisely to the E-L equations \eqref{cons-law}. The
same set of equations may be obtained as a geometric condition for
the immersion of the surfaces $X_k$ in $\mathbb{R}^{N^2-1}$, i.e.
as Gauss-Mainardi-Codazzi equations for the surfaces given by
\eqref{X}.

An explicit solution, vanishing at complex infinity was found for
equations \eqref{spectral} in \cite{DHZ}:
 \be\label{PhifromP}
\Phi_k=\mathbb{I}+\frac{4\la}{(1-\la)^2}\sum\limits_{j=0}^{k-1}P_j-\frac{2}{1-\la}P_k,
\ee\be
{\Phi_k}^{-1}=\mathbb{I}-\frac{4\la}{(1+\la)^2}\sum\limits_{j=0}^{k-1}P_j-\frac{2}{1+\la}P_k.
\ee
Similarly the integration \eqref{X} has explicitly been carried
out for the $\cpn$ models defined on $S^2$ and having finite
action \cite{GY1}. By choosing the integration constant so that
the $X_k$ are traceless, we obtain the following solution
\be
\label{XfromP}
X_k=-i\left(P_k+2\sum\limits_{j=0}^{k-1}P_j\right)+\frac{i(1+2k)}{N}\mathbb{I},
\qquad k=0,1,\ldots,N-2.
\ee
The above formula will be referred to as the GY formula. Finally the
Sym-Tafel formula \cite{Sym1,Sym2,Sym3,Taf} yields $X_k$ from
$\Phi_k$ as
\be\label{ST}
X_k=\alpha(\la)\Phi_k^{-1}\p_{\la}\Phi_k+\frac{(1+2k)}{N}\mathbb{I},
\qquad k=0,1,\ldots,N-2.
\ee
For the $\cpn$ models, $\alpha(\la)$ was found to be equal to
$2/(1-\la^2).$ We have found another way of obtaining $X_k$ from
$\Phi_k$, from its asymptote at large values of the spectral
parameter $\la$, namely
\be\label{XbyLimit}
X_k=i\,\frac{2k+1}{N}\,\mathbb{I}
+\frac{i}{2}\lim\limits_{\la\to\infty}\left[\la(\mathbb{I}-\Phi_k)\right].
\ee
However this procedure is obviously limited to the $\cpn$ models
while the Sym-Tafel formula is universal. ~

The solutions $z=f/|f|$ have well-known physical interpretation as
localized soliton-like objects, i.e. instantons. As a rule the
holomorphic solution ($k=0$) is recognized as instanton, the
antiholomorphic one ($k=N-1$) as antiinstanton, while the
intermediate solutions ($k=1,...,N-2$, possible in the $\cpn$ models
for $N\ge 2$) describe various mixed instanton-antiinstanton states.

\section{Recurrence in the projection space}
The recurrence in the projection space is a construction of new
projectors in terms of the previous ones. First we look for an
operator which transforms each projector $P_i$ to the next one
$P_{i-1} (0\leq i\leq N-2)$. Each of the projectors maps onto a
one-dimensional space and altogether they constitute a partition of
the identity matrix. This way we may systematically build consecutive
dimensions in the partition of unity, starting from a holomorphic or
antiholomorphic solution of the Euler-Lagrange equations
\eqref{cons-law}.

Let $\mathbf{\Pi_\pm}$ be operators acting on those projectors in the
way
\be
\label{Ppm}
\mathbf{\Pi_-}(P_k)=P_{k-1}, \qquad \mathbf{\Pi_+}(P_k)=P_{k+1}.
\ee
These operators play the role of annihilation and creation
operators (respectively) in the space  of projectors. However they
are nonlinear and the objects on which they act have to remain
normalized to retain their projective character. For this reason
they cannot be used to construct the ``particle number operator''.

It is proven in Appendix A that the operators \eqref{Ppm} may be cast
into the forms
\be
\label{Pi-} \mathbf{\Pi_-}(P)=\frac{\bp P\c P\c \p P}{\tr(\bp P\c P
\c\p P)}=\frac{(\mathbb{I}-P)\c \bp P\c \p P}{\tr(\bp P\c P\c\p
P)}=\frac{\bp P\c \p P\c (\mathbb{I}-P)}{\tr(\bp P\c P\c \p P)},
\ee
and
\be
\label{Pi+} \mathbf{\Pi_+}(P)=\frac{\p P\c P \c\bp P}{\tr(\p P\c P
\c\bp P)}=\frac{(\mathbb{I}-P)\c \p P\c \bp P}{\tr(\p P\c P\c\bp
P)}=\frac{\p P\c \bp P\c (\mathbb{I}-P)}{\tr(\p P\c P\c \bp P)},
\ee
where the traces in the denominators are different from zero
unless the whole matrix is zero (which occurs when applying
$\mathbf{\Pi_-}$ to the holomorphic or $\mathbf{\Pi_+}$ to the
antiholomorphic solution).
At the end of Appendix A we prove that the resulting matrices
$\mathbf{\Pi_-}(P)$ and $\mathbf{\Pi_+}(P)$ have the orthogonal
projective property $M^2=M$ and $M^{\dagger}=M$, provided that the
argument $P$ is a projector mapping onto a one-dimensional
subspace. Non-vanishing of the traces in \eqref{Pi-} and
\eqref{Pi+} is a by-product of the proof (see the comment to
\eqref{tr}).
\begin{exam}[Action of $\Pi_{\pm}$ in ${\C P^2}$]

A projector corresponding to the holomorphic Veronese solution of
the Euler-Lagrange equations \eqref{E-L}, which itself is a
solution of the \eqref{cons-law}  reads {\rm \cite{WZ1}}
\be
P_0=-\frac{2}{\left(|\xi|^2+1\right)^3} \,\left(
\begin{array}{lll}
 1 & \sqrt{2} \bar{\xi} & \bar{\xi}^2 \\
 \sqrt{2} \xi  & 2 |\xi|^2 & \sqrt{2} |\xi|^2 \bar{\xi}
   \\
 \xi ^2 & \sqrt{2} |\xi|^2 \xi  & |\xi|^4
\end{array}
\right)
\ee
An action of the ``creation operator'' $\mathbf{\Pi_+}$ \eqref{Pi+}
converts it into a projector corresponding to a mixed solution
\be
P_1=\frac{1}{\left(|\xi|^2+1\right)^2}\,\left(
\begin{array}{lll}
 2 |\xi|^2 & \sqrt{2} \left(|\xi|^2-1\right) \bar{\xi} &
   -2 \bar{\xi}^2 \\
 \sqrt{2} \left(|\xi|^2-1\right) \xi  &
   \left(|\xi|^2-1\right)^2 & -\sqrt{2} \left(|\xi|^2-1\right)
   \bar{\xi} \\
 -2 \xi ^2 & -\sqrt{2} \left(|\xi|^2-1\right) \xi  & 2 |\xi|^2
\end{array}
\right)
\ee
This procedure can be repeated once to yield a projector mapping
on the direction of the antiholomorphic solution of \eqref{E-L}.
Further application of the creation operator, i.e. on the
antiholomorphic projector yields an indeterminate expression of
the form $0/0$, since an action of the $\p$ operator on an
antiholomorphic function yields zero both in the numerator and the
denominator of \eqref{Pi+}. .
\end{exam}
These operators will further be used to construct the corresponding
``creation'' and ``annihilation'' operators for the wave functions
$\Phi_k$ and for the immersion functions $X_k$.

The corresponding recurrence relations for the wave functions
$\Phi_k$ may be obtained from the solution of the spectral problem
(\ref{spectral}). The relations are more conveniently expressed in
terms of an auxiliary function
\be
\Psi_k=(1-\lambda)^2(\mathbb{I}-\Phi_k).
\ee
As in the case of the projection matrices $P_k$, the
``creation/annihilation'' operators $\mathbf{\Lambda}_\pm$  raise
or lower the index of $\Psi_k$ by one. The operators, which depend
on the spectral parameter $\la$, read
\be
\label{lambda-}
\mathbf{\Lambda_-}\left(\Psi(\lambda)\right)=\frac{1}{2}[(1+\la)\Psi(\la)
-(1-\la)\Psi(-\la)]+2(1+\la)\mathbf{\Pi_-}\left(\frac{1}{4}[\Psi(\la)+\Psi(-\la)]\right),
\ee
and
\be
\label{lambda+}
\mathbf{\Lambda_+}\left(\Psi(\lambda)\right)=\frac{1}{2}[(1-\la)\Psi(\la)+(1+\la)\Psi(-\la)]
+2(1-\la)\mathbf{\Pi_+}\left(\frac{1}{4}[\Psi(\la)+\Psi(-\la)]\right).
\ee
where $\Psi(-\la)$ may also be expressed in terms of $\Psi(\la)$ if
we make use of the symmetry $\Phi^{-1}(\la)=\Phi(-\la)$, namely
\be
\Psi(-\la)=-(1+\la)^2\Psi(\la)[(1-\la)^2\mathbb{I}-\Psi(\la)]^{-1}.
\ee
A simple proof of these formulae may be found in Appendix B.
\begin{exam}[Action of $\Lambda_\pm$ in ${\C P^3}$]

For the ${\C P^3}$ model the wave function for the spectral
problem whose compatibility condition is \eqref{cons-law} may be
constructed according to \eqref{PhifromP}. If we use that equation
with $k=0$, we obtain the wave function $\phi_0$. The auxiliary
function $\Psi_0$ may be obtained from it as
$(1-\lambda)^2(\mathbb{I}-\Phi_0)$. It reads
\be
\Psi_0=\frac{2 (1-\lambda)}{\left(|\xi|^2+1\right)^3}\,
 \left(
\begin{array}{llll}
 1 & \sqrt{3} \bar{\xi} & \sqrt{3}
\bar{\xi}^2 & \bar{\xi}^3 \\
 \sqrt{3} \xi  & 3 |\xi|^2
    &
   3 |\xi|^2
\bar{\xi} & \sqrt{3} |\xi|^2
   \bar{\xi} \\
 \sqrt{3} \xi ^2 & 3 |\xi|^2
   \xi  & 3 |\xi|^4
    &
   \sqrt{3} |\xi|^4 \bar{\xi} \\
 \xi ^3 & \sqrt{3} |\xi|^2 \xi ^2 & \sqrt{3} |\xi|^4 \xi  &
   |\xi|^6
\end{array}
\right).
\ee
An action of the $\Lambda_+$ on $\Psi_0$ yields the next $\Psi$
i.e. $\Psi_1=(1-\lambda)^2(\mathbb{I}-\Phi_1)$, where $\Phi_1$ is
another wave function, whose spectral problem \eqref{spectral}
yields the equation \eqref{cons-law} as compatibility condition,
with $P_1$ instead of $P_0$. The new wave function $\Phi_k$ may
also be constructed in terms of the projectors according to
\eqref{PhifromP} with $k=1$. The new $\Psi_1$ has the form
\begin{eqnarray}
\!\!\!\!\!\!\!\!\!\!\!\!\!\!\!\!\!\!\!\!\!\!\!\!\!\!\!\!\!\!\!\!\Psi_1=-\frac{2}{\left(|\xi|^2+1\right)^3}\,
\left(
\begin{array}{llll}
 3 (\lambda -1) |\xi|^2+2 \lambda  & \sqrt{3} \left[2 (\lambda -1)
|\xi|^2+\lambda +1\right] \bar{\xi} \\
 \sqrt{3} \left[2 (\lambda -1) |\xi|^2+\lambda +1\right] \xi  & 4 (\lambda
-1) |\xi|^4+2 (\lambda +2) |\xi|^2+\lambda -1 \\
 \sqrt{3} \left[(\lambda -1) |\xi|^2+2\right] \xi ^2 & \left[2 (\lambda -1)
|\xi|^4+(\lambda +5) |\xi|^2+2 (\lambda -1)\right] \xi  \\
 (3-\lambda) \xi ^3 & \sqrt{3} \left(2 |\xi|^2+\lambda -1\right) \xi ^2
\end{array}\right.&&\nn\\
\!\!\!\!\!\!\!\!\!\!\!\!\!\!\!\!\!\!\!\left.\begin{array}{llll}
 \sqrt{3} \left[(\lambda -1)
   |\xi|^2+2\right] \bar{\xi}^2 & (3-\lambda) \bar{\xi}^3 \\
   \left[2 (\lambda
   -1) |\xi|^4+(\lambda +5) |\xi|^2+2 (\lambda -1)\right] \bar{\xi} & \sqrt{3}
\left(2 |\xi|^2+\lambda -1\right) \bar{\xi}^2 \\
 |\xi|^2
   \left[(\lambda -1) |\xi|^4+2 (\lambda +2) |\xi|^2+4 (\lambda -1)\right] &
\sqrt{3} |\xi|^2 \left[(\lambda +1) |\xi|^2+2 (\lambda -1)\right]
   \bar{\xi} \\
   \sqrt{3} |\xi|^2 \left[(\lambda +1) |\xi|^2+2 (\lambda -1)\right]
   \xi  & |\xi|^4 \left[\left(2 |\xi|^2+3\right) \lambda -3\right]
\end{array}
\right)\!\! .&&
\end{eqnarray}
 Such an action of the nonlinear
operator $\Lambda_+$ may be repeated by applying it consecutively to
$\Psi_1$ and $\Psi_2=\Lambda_+(\Psi_1)$. Further application of the
operator yields a trivial result. Inversely, we can go down the
ladder by applying $\Lambda_-$ to $\phi_3,~\phi_2~\text{and}~\phi_1$.
\end{exam}
Although the usual creation and annihilation operators have well
defined interpretation for wave functions, our nonlinear operators
cannot be interpreted that way.

Finally the recurrence relations may be constructed for the immersion
functions $X_k$. In this case, the value of the index $k$ appears in
the formulae explicitly. Note that in principle the explicit use of
$k$ can be eliminated from \eqref{Pk from Xk} by expressing $k$ in
terms of $\tr(X^2)$, which is equal to $(2k+1)^2/N-(4k+1)$. However
this does not make much sense as the immersion functions $X_k$ are
only well defined for $k=0,...,N-1$.

It follows from \eqref{XfromP} that the projectors $P_k$ may be expressed as
\be
\label{Pk from Xk}
P_k={X_k}^2-2i\left(\frac{2k+1}{N}-1\right)X_k-\frac{2k+1}{N}
\left(\frac{2k+1}{N}-2\right)\mathbb{I},
\ee
which allows us to write the ``annihilation'' operator as
\be
\label{chi-}
{\bm \chi}_-(X_k)=X_k+i[\mathbf{\Pi_-}(P_k)+P_k]-(2i/N)\mathbb{I},
\ee
where $P_k$ are given by \eqref{Pk from Xk}. Similarly, the
``creation'' operator may be defined by
\be
\label{chi+}
{\bm \chi}_+(X_k)=X_k-i[\mathbf{\Pi_+}(P_k)+P_k]+(2i/N)\mathbb{I}.
\ee
\begin{exam}[Action of $\chi_\pm$ in ${\C P^2}$, this time ``descending''
the ladder]

The surface $X_1$, whose condition of immersion in $\mathbb{R}^8$ is
the equation \eqref{cons-law} for $P=P_1$, may be written in the
matrix form as
\be\label{X1}
X_1= i \left(
\begin{array}{lll}
 1 & 0 & 0 \\
 0 & 1 & 0 \\
 0 & 0 & 1
\end{array}
\right)-\frac{1}{\left(k^2+1\right)^2}\, \left(
\begin{array}{lll}
 2 i & i \sqrt{2} \bar{\xi} & 0 \\
 i \sqrt{2} \xi  & i \left(|\xi|^2+1\right) & i
   \sqrt{2} \bar{\xi} \\
 0 & i \sqrt{2} \xi  & 2 i |\xi|^2
\end{array}
\right).
\ee
If we apply the operation $\chi_-$ to \eqref{X1}, then we obtain a
matrix (after some simplification)
\be
X_0=\frac{1}
   {3} i \left(
\begin{array}{lll}
 1 & 0 & 0 \\
 0 & 1 & 0 \\
 0 & 0 & 1
\end{array}
\right)-\frac{1}{\left(|\xi|^2+1\right)^2}\, \left(
\begin{array}{lll}
 i & i \sqrt{2} \bar{\xi} & i \bar{\xi}^2
   \\
 i \sqrt{2} \xi  & 2 i |\xi|^2 & i \sqrt{2} |\xi|^2
   \bar{\xi} \\
 i \xi ^2 & i \sqrt{2} |\xi|^2 \xi  & i |\xi|^4
\end{array}
\right).
\ee
This is the matrix form of a two-dimensional surface immersed in
$\mathbb{R}^8$ representing the soliton surface whose condition for
immersion in $\mathbb{R}^8$ (the Gauss-Mainardi-Codazzi equations) is
\eqref{cons-law} for the projector $P=P_0$.

 The surface $X_1$ may in turn be obtained by similar procedure performed on the surface
 $X_2$, whose condition for immersion is \eqref{cons-law} for the
 projector $P_2$ which maps on the direction of the antiholomorphic
 solution of \eqref{E-L}.
 \end{exam}
We may also express each of the projection operators $P_k$ as a
linear function of the surfaces. However, this requires knowledge
of $X_0,...,X_{k-1}$, thus making the recurrence deeper. Namely,
from the equations of the surfaces in terms of the projectors
\eqref{XfromP} we obtain
\be
\label{PkX} P_k=i\sum\limits_{j=1}^k
(-1)^{k-j}\left(X_j-X_{j-1}\right)+(-1)^k i
X_0+\frac{1}{N}\mathbb{I},
\ee
which may be used to construct the recurrence relations involving all
the lower-index $X_j,~j=0,...k-1$. In a similar way a downwards
recurrence might be obtained, involving all the higher-index $X_j$.

Equation \eqref{Pk from Xk} directly follows from the equations
\eqref{XfromP}. A short derivation of that equation is given in
Appendix C.

\section{Geometrical aspects of the $\cpn$ model}

Let us now explore certain geometrical characteristics of surfaces
immersed in the $\su{N}$ algebra and express them in terms of the
projectors $P_k$. These geometrical properties include the Gaussian
curvature, the mean curvature vector, the topological charge, the
Willmore functional and the Euler-Poincar\'{e} character (see e.g.
\cite{MS,NS,WZ}). Under the assumption that the $\cpn$ model is
defined on the Riemann sphere $S^2$ and the associated action
functional of this model is finite we can show that surfaces are
conformally parametrized. The proof is similar to that given in
\cite{GY2}. In Appendix D we demonstrate that whenever the equations
of motion (\ref{cons-law}) are satisfied, the holomorphic quantity
\be
\label{current}
\begin{split}
J_k=&(g_k)_{11}=-\frac{1}{2}\tr(\p
X_k)^2=\frac{1}{(P_k)_{11}}\left[(\p P_k)^2\cdot
P_k\right]_{11}=\frac{1}{(P_k)_{11}}\left[P_k\cdot(\p
P_k)^2\right]_{11},\\ &\bp J_k=0
\end{split}
\ee
and its respective complex conjugate
\be
\label{currentb}
\begin{split}
\bar{J}_k=&(g_k)_{22}=-\frac{1}{2}\tr(\bp
X_k)^2=\frac{1}{(P_k)_{11}}\left[(\bp P_k)^2\cdot
P_k\right]_{11}=\frac{1}{(P_k)_{11}}\left[P_k\cdot(\bp
P_k)^2\right]_{11},\\ &\p \bar{J}_k=0
\end{split}
\ee
vanish in the first fundamental form of surfaces $X_k$ associated
with (\ref{XfromP}). The first fundamental form $I_k$ becomes
\be
\label{first}
I_k=2(g_k)_{12}d\xi d\bar{\xi},
\ee
where the nonzero component of the induced metric $(g_k)_{ij}$ on
surfaces $X_k$ are given by
\be
\label{first2} (g_k)_{12}=-\frac{1}{2}\tr{(\p X_k\cdot\bp
X_k)}=\frac{1}{(P_k)_{11}}\left(\bp P_k\cdot\p P_k\cdot
P_k\right)_{11}=\frac{1}{(P_k)_{11}}\left(P_k\cdot\bp P_k\cdot\p
P_k\right)_{11}.
\ee
Here the index inside the parentheses in $g$ refers to the number
of the surface, while the other two indices denote the appropriate
components of the metric tensor.

It follows from the Bonnet theorem that the surfaces $X_k$ are
determined uniquely up to Euclidean motions by their first
fundamental forms (\ref{first}) and their second fundamental forms
\be
\label{second} II_k=(\p^2X_k-(\Gamma_k)^1_{11}\p X_k)d\xi^2+2\p\bp
X_k d\xi d\bar{\xi}+(\bp^2X_k-(\Gamma_k)^2_{22}\bp X_k)d\bar{\xi}^2,
\ee
where the immersion function $X_k$ is expressed in terms of
projectors $P_k$ by the formula (\ref{XfromP}) and the nonzero
Christoffel symbols of the second kind are given by
\be
\label{second2} (\Gamma_k)^1_{11}=\p\ln{(g_k)_{12}},\qquad
(\Gamma_k)^2_{22}=\bp\ln{(g_k)_{12}}.
\ee
Since the projectors $P_0,\ldots,P_{N-1}$ are uniquely determined by
the surfaces $X_k$ (see (\ref{Pk from Xk}) and (\ref{PkX})) it
follows that the projectors $P_k$ are determined (to that extent) by
the fundamental forms (\ref{first}) and (\ref{second}). When $J_k=0$,
the Gaussian curvatures $\mathcal{K}_k$ and the mean curvature vector
$\mathcal{H}_k$ (written as a matrix) take the simple form
\be
\label{gauss}
\mathcal{K}_k=\frac{-1}{(g_k)_{12}}\p\bp\ln{(g_k)_{12}},
\ee
and
\be
\label{gauss2}
\mathcal{H}_k=\frac{2}{(g_k)_{12}}\p\bp X_k,
\ee
respectively, where the immersion function $X_k$ is given by (\ref{XfromP}).
\begin{exam}[Geometrical properties of the family generated by the Veronese solutions
in  ${\C P^2}$]

We may easily determine the geometrical characteristics for the
holomorphic Veronese solution of \eqref{E-L} or the corresponding
projector solutions of \eqref{cons-law} and for the solutions
obtained from them by application of the ``creation'' operator
\eqref{Pi+}. The first fundamental form is completely determined by
\be\label{g12}
(g_0)_{12} = \frac{1}{\left(|\xi|^2+1\right)^2}=(g_2)_{12}, \qquad
(g_1)_{12}=\frac{2}{\left(|\xi|^2+1\right)^2},
\ee
where the index inside the parentheses in $g$ is $0$ for the
holomorphic, $1$ for the mixed solution and $2$ for the
antiholomorphic solution.

The second fundamental form is determined by the Christoffel
symbols. They have the same values for all three surfaces as the
constant factor 2 in \eqref{g12} does not influence the
logarithmic derivative in \eqref{second2}.

The nonzero Christoffel symbols read (with the same convention about
the indices)
\be
(\Gamma_0)^1_{11}=(\Gamma_1)^1_{11}=(\Gamma_2)^1_{11}-\frac{2\bar{\xi}}{|\xi|^2+1},\qquad
(\Gamma_0)^2_{22}=(\Gamma_1)^2_{22}=(\Gamma_2)^2_{22}=-\frac{2 \xi}
{|\xi|^2+1}.
\ee
The Gaussian curvature may be obtained in a straightforward way from
\eqref{gauss} as
\be
\mathcal{K}_0=\mathcal{K}_2=2,\qquad \mathcal{K}_1=1.
\ee
Hence all these surfaces have constant positive Gaussian curvature.

The mean curvature is given by a rather complicated traceless matrix
expression (or a vector expression if we decompose the matrix in the
basis of the Gelfand matrices). In the matrix form we get e.g. for
the surface corresponding to the holomorphic solution
\be
\mathcal{H}_0=\frac{4 i}{(|\xi|^2+1)^2}\left(\begin{array}{lll}1-
2|\xi|^2&\sqrt{2}(2-|\xi|^2)\bar{\xi}& 3\bar{\xi}^2\\
\sqrt{2}(2-|\xi|^2)\xi
&-(|\xi|^4-4|\xi|^2+1)&\sqrt{2}(2|\xi|^2-1)\bar{\xi}\\
3\xi^2&\sqrt{2}(2|\xi|^2-1)\xi&|\xi|^2(|\xi|^2-2)\end{array}\right).
\ee
However the mean curvature proves to be a vector of constant norm,
namely square of the norms calculated according to \eqref{scalar}
are
\be\label{Hnorm}
(\mathcal{H}_0,\mathcal{H}_0)=4\,(\mathcal{H}_1,\mathcal{H}_1)=
(\mathcal{H}_2,\mathcal{H}_2)=16.
\ee
\end{exam}
This result may be used to calculate the Willmore functional. The
Willmore functional (also called the total squared mean curvature
vector) is defined by
\be\label{willmore-def}
W_k=\frac{1}{4}\int\!\!\!\int_{\Omega}||\mathcal{H}_k||^2\sqrt{|\mathrm{det}
(g_k)_{ij}|} d\xi^1 d\xi^2,
\ee
where $\Omega\in\mathbb{C}$ is an open connected and simply connected
set, while the norm is defined in terms of the scalar product in the
usual way: $||\cdot||=(\cdot,\cdot)^{1/2}$. Hence, in our example, if
take $\Omega=S^2$ in \eqref{willmore-def} (i.e. we integrate over the
whole Riemann sphere), we obtain from \eqref{Hnorm} and \eqref{g12}
\be\label{willmore}
W_0 = 4 W_1 = W_2 =4\pi
\ee

In a similar way we may calculate a few other global characteristics
of the soliton surfaces defined by the immersion functions $X_k$. In
particular a significant quantity which characterizes solutions
satisfying the $\cpn$ model equations (\ref{cons-law}) is the
topological charge associated with the surfaces $X_k$ \cite{Din}
\be\label{top1}
Q_k= \frac{1}{\pi}\int\!\!\!\int_{S^2}\p\bar{\p}\ln |f_k|^2 d\xi^1
d\xi^2,
\ee
which may be transformed into
\be
\label{top} Q_k=-\frac{1}{\pi}\int\!\!\!\int_{S^2}\tr{(P_k\cdot[\p
P_k,\bp P_k])}d\xi^1 d\xi^2.
\ee
The integral (\ref{top}) exists and it is a topological invariant of
the surfaces given by (\ref{chi-}) or (\ref{chi+}). It is an integer
which globally characterizes the  surfaces $X_k$.

In the case of compact oriented and connected surfaces $X_k$ another
topological invariant: the Euler-Poincar\'{e} characteristic is given
by
\be
\label{EP}
\Delta_k=\frac{i}{2\pi}\int\!\!\!\int_{S^2}\mathcal{K}_k(g_k)_{12}d\xi
d\bar{\xi}=-\frac{i}{2\pi}\int\!\!\!\int_{S^2}\p\bp\ln{(g_k)_{12}}d\xi
d\bar{\xi}=-\frac{1}{\pi} \int\!\!\!\int_{S^2}\p\bp\ln{(g_k)_{12}}
d\xi^1 d\xi^2.
\ee
If we know the projector $P_k$ explicitly, the calculation of
$\Delta_k$ is straightforward.
\begin{exam}

In the particular case where $N=3$ (the $\mathbb{C}P^2$ model)
\eqref{top} and \eqref{EP} turn into
\be\label{Q}
Q_0=2,\qquad Q_1=0, \qquad Q_2=-2
\ee
for the topological charges, and
\be
\Delta_0=\Delta_1=\Delta_2=2
\ee
for the Euler-Poincar\'e characteristics. \\
The result \eqref{Q} is in accordance with the values of the
topological charge obtained in {\rm \cite{Din}}. The value of the
topological charge distinguishes the instantons ($Q=2$ for a
one-instanton state in $\mathbb{C}P^2$, $Q=-2$ for the antiinstanton
state, which produces the same winding over the target sphere but in
the opposite direction).
\end{exam}

\section{Singularities of the $\cpn$ model}
In what follows we do not impose the assumption that the action
functional of the $\cpn$ model \eqref{lagr-z} is finite.

The E-L equations of the $\cpn$ model \eqref{E-L} are autonomous,
hence they do not have fixed singularities at finite points. On the
other hand, as nonlinear equations, they might in principle have
movable singularities. Let us limit ourselves to solutions without
branch points or essential singularities. The scaling invariance puts
limits to the singular behavior of such solutions: the singularities
disappear in the invariant description.  The following statements
directly follow from the scaling invariance:
\begin{enumerate}
 \item
If a $j$-th component of a homogeneous field coordinate $f$ has a
pole of order $p$ greater than or equal to the order of other poles
at a point $\xi_0$, then the solution may be multiplied by
$(\xi-\xi_0)^p$. This yields a solution $f$ of the E-L equations
which constitutes the same solution in the invariant variables. An
appropriate multiplication by a product of such factors can always be
performed if the number of poles is finite. Also in many cases with
an infinite number of poles we can build a holomorphic function which
would regularize the solution, making use of the Weierstrass theorem
(provided that the poles have no finite accumulation point). This
multiplication makes the solution regular and we will refer to the
procedure as regularization.
\item
The regularization of a field coordinate $f$ through the
multiplication by a singularity-removing factor $(\xi-\xi_0)^p$ would
introduce zeros at the point $\xi_0$ in all the components which were
regular or had poles of lower order than $p$. In such a case the
usual normalization of $f$ by setting its first component to one may
be impossible.
\item
In the $\cpn$ model we also consider functions which are not
holomorphic as the E-L equations \eqref{E-L} depend on both $f$
and $\bar{f}$. To perform a singularity analysis in such cases,
both independent variables are extended to separate complex planes
and the field coordinates $f$ intrinsically become functions of
two complex variables. However all the previous and further
considerations hold, with the modification that $\xi-\xi_0$ is
replaced by some function $F(\xi,\bar{\xi})$ which would vanish at
the line of singularity (except that the class of exceptions is
richer in two dimensions than in one).
\item
To summarize, the regularization leaves invariant
\begin{itemize}
 \item
the E-L equations in both forms \eqref{E-L}, \eqref{cons-law} and the
action functional \eqref{action};
\item
the projectors $P_k,~k=0,...,N-1$ as well as any projection operators
in the algebra $\su{N}$ of anti-Hermitian matrices (or $i\su{N}$ in
the case of Hermitian matrices);
\item
the surfaces $X_k$ with all their induced metrics $(g_k)_{ij}$ and
curvature properties $\mathcal{K}_k$ and $\mathcal{H}_k$;
\item
the ``creation'' and ``annihilation'' operators
$\mathbf{\Pi}_{\pm},~\mathbf{\Lambda}_{\pm}$ and ${\bm \chi}_{\pm}$.
\item
The classical operators $P_-$ and $P_+$ are covariant in the sense
that $P_\pm\left(f_k(\xi-\xi_0)^p\right)=(\xi-\xi_0)^p
P_\pm\left(f_k\right)$, which allows for regularization of the
Din-Zakrzewski procedure \cite{Din}.
\end{itemize}
\end{enumerate}

\section{Summary and concluding remarks}

The objective of this paper was to provide an invariant description
of recurrence relations for the completely integrable $\cpn$ sigma
models defined on the Riemann sphere $S^2$ when its action functional
is finite. We have determined the connection between successive
projector operators, wave functions of the linear spectral problem
and immersion functions which immerse the surfaces in the $\su{N}$
algebra in such a way that they preserve conformal invariance.
Through this link, we found explicit expressions for these quantities
and established a commutative diagram for them. An advantage of the
presented approach is that, without reference to any additional
consideration, the recurrence relations give a very useful tool for
constructing each successive surface associated with the $\cpn$ sigma
model from the knowledge of the previous one. We have also analyzed
the asymptotic properties of solutions of the $\cpn$ model in
neighborhoods of zeros and poles (excluding branch points and
essential singularities) and demonstrated that the singularity
structures of meromorphic solutions of the model do not influence the
above-mentioned invariant quantities. Consequently, we have shown
that the surfaces associated with the $\cpn$ model are regular.
Furthermore, we provide a certain geometrical setting which allows us
to obtain explicit formulae in terms of the projector $P_k$ for the
Gaussian and mean curvatures, the Willmore functional, the
Euler-Poincar\'{e} character and the topological charge of the
considered surfaces. This allows us to study certain global
properties of the surfaces as illustrated by concrete examples of
surfaces associated with the $\mathbb{C}P^2$ and $\mathbb{C}P^3$
models. In particular we have shown that for the Veronese vectors we
obtain constant positive Gaussian curvatures as expected.

It may be worthwhile to extend the investigation of surfaces to
the case of the sigma models defined on other homogeneous spaces
via Grassmannian models. This case can lead to different classes
and more diverse types of surfaces than those investigated in this
paper, including those with constant negative Gaussian curvature.
These types of surfaces immersed in Lie algebras are known to have
many fundamental applications in physics, chemistry and biology
(see e.g. \cite{Pol,Dav,NPW,Ou,Saf,Lan}).  This task will be
undertaken in a future work.

{\bf Acknowledgments}
A.M.G.'s work was supported by a research grant from NSERC of Canada.
P.G. wishes to acknowledge and thank the Mathematical Physics
Laboratory of the Centre de Recherches Math\'{e}matiques for their
hospitality during his visit to the Universit{\'e} de Montr{\'e}al.

\appendix

\section{Derivations of the recurrence relations for the projection operators \eqref{Pi-}
and \eqref{Pi+}}

To construct the recursion operator, we start with the $P_\pm$
operators \eqref{Ppm}, which raise or lower the index of the
homogeneous field coordinates $f_k$ by one.

The $k$-th coordinate $f_k$ may be regained from the respective
projector $P_k$ by an extraction of its first column
\be
\label{P2f}
f_k = \frac{1}{(P_k)_{11}} \,P_k\cdot\left(\begin{array}{c}
 1\\ 0\\ \vdots\\ 0
\end{array}\right),
\ee
where $(P_k)_{11}$ is the 1st row-1st column element of the matrix
$P_k$. The first row of its Hermitian conjugate is obtained
similarly by multiplying on the left by $(1,~0,...,0)$.

This equation yields $f_k$ with the 1st component of $f_k$ normalized
to one. For the sake of simplicity the derivation will be done for
that case. If $(P_k)_{11}=0$ the first component of $f_k$ is
zero. In that case we can get the $f_k$ by extracting another column
of $P_k$, which is done by multiplying with a vector having $1$ at
the other position (and zeros elsewhere).

Substituting \eqref{P2f}, together with its Hermitian conjugate, into
\eqref{P-} and \eqref{P+}, we obtain the nonlinear ``creation
operator'' $\mathbf{\Pi_+}$ for the projectors $P_k$
\be\label{P-oper}
\mathbf{\Pi_+}(P)=\frac{(\mathbb{I}-P)\cdot\p P\cdot \mathbb{I}_0
\cdot
\bar{\p}P\cdot(\mathbb{I}-P)}{\left[\bar{\p}P\cdot(\mathbb{I}-P)\cdot\p
P\right]_{11}},
\ee
where $[~]_{11}$ denotes the leftmost-uppermost element of the matrix
while
\be
\mathbb{I}_0=\left(\begin{array}{ccc}1&\hdots & 0\\ 0 & \hdots & 0\\
\hdots&\hdots&\hdots \\ 0 & \hdots & 0 \end{array}\right).
\ee
In the transition from \eqref{P2f} to \eqref{P-oper} we used the
scaling invariance to get rid of the factor $(P_k)_{11}$.

This operator may further be simplified if we use the following
property of projectors
\be\label{proj}
(\mathbb{I}-P)\c\p P=(\p P)\c P, \qquad  (\p P)\c (\mathbb{I}-P)=P\c \p P.
\ee
The identity \eqref{proj} yields equation \eqref{Pi+} in a
straightforward way if we note that any Hermitian projection operator
$P$ mapping onto a one-dimensional space and satisfying $\tr(P)=1$
may be represented as $U^{-1}\mathbb{I}_0 U$, where $U$ is a unitary
matrix (the diagonalized $P$ has only one nonzero element, equal $1$
and this $1$ may always be placed at the upper left corner as in the
$\mathbb{I}_0$ matrix). Moreover by direct calculation
\be\label{P=U^2}
P_{11}=(U^{-1})_{11}U_{11}={U_{11}}^2
\ee
as $U^{-1}=U^\dagger$. Hence
\be
\label{unitary} P\c\mathbb{I}_0\c P=U^{-1}\c \mathbb{I}_0\c
U\c\mathbb{I}_0\c \mathbb{I}_0 \c U^{-1}\mathbb{I}_0\c U = P_{11}\,P.
\ee
as we have we have
\be\label{I0I0}
\mathbb{I}_0\cdot M\cdot \mathbb{I}_0=M_{11}\mathbb{I_0}
\ee
 for any
matrix $M$. Equation \eqref{unitary} yields the numerator of
\eqref{Pi+} up to a constant factor. The denominator immediately
follows from the normalization $\tr(\mathbf{\Pi_+}(P))=\tr(P)=1$,
provided that the matrix is nonzero (see below for the proof that its
trace is also nonzero).

The ``annihilation'' operator $\mathbf{\Pi_-}$ is obtained from
$\mathbf{\Pi_+}$ by exchanging the partial derivatives
$\p\rightleftharpoons\bp$.

The projective property of the resulting operators
$\mathbf{\Pi_-}(P)$ and $\mathbf{\Pi_+}$ (\ref{Pi-}, \ref{Pi+}) may
be proven by means of the same unitary conversion of $P$. Let us
check the square of  $\mathbf{\Pi_+}(P)$ \eqref{Pi+}:
\be\label{Pi+Pi+}
\mathbf{\Pi_+}(P)\c \mathbf{\Pi_+}(P)=\frac{\p P\c P\c \bp P\c \p P\c
P\c \bp P}{[\tr(\p P\c P\c\bp P)]^2} =\frac{\p P\c U^{-1}\c
\mathbb{I}_0\c U \c\bp P\c \p P\c U^{-1}\c \mathbb{I}_0\c U  \c \bp
P}{[\tr(\p P\c P\c\bp P)]^2}.
\ee
If the numerator of\eqref{Pi+Pi+} is a zero matrix, then the
projective property is trivial. If the numerator is a nonzero matrix,
then, according to \eqref{I0I0} its central part which begins and
ends with $\mathbb{I}_0$ is a diagonal matrix with only one nonzero
element in the top left position. Hence it is equal to its trace
multiplied by $\mathbb{I}_0$:
\begin{eqnarray}\label{tr}
\mathbb{I}_0\c U \c\bp P\c \p P\c U^{-1}\c
\mathbb{I}_0=\tr\left(\mathbb{I}_0\c U \c\bp P\c \p P\c U^{-1}\c
\mathbb{I}_0\right)\mathbb{I}_0&&\nn\\
\!\!\!\!=\tr(U^{-1}\c\mathbb{I}_0\c U\c\bp P\c \p P)\mathbb{I}_0
=\tr(P\c \bp P\c \p P)\mathbb{I}_0=\tr(\p P\c P\c\bp
P)\mathbb{I}_0.&&
\end{eqnarray}
It follows from \eqref{tr} that the trace of $\p P\c P\c\bp P$ is
nonzero whenever the matrix is nonzero. Otherwise the matrix would be
nilpotent, but this is impossible for a nonzero Hermitian matrix.

Combining \eqref{tr} with the rest of the equation \eqref{Pi+Pi+}
we eventually obtain
\be
\mathbf{\Pi_+}(P)\c \mathbf{\Pi_+}(P)=\tr(\p P\c P\c\bp P)\,\p P\c
U^{-1}\c \mathbb{I}_0\c U \c\bp P/[\tr(\p P\c P\c\bp
P)]^2=\mathbf{\Pi_+}(P).
\ee
The same property obviously holds for $\mathbf{\Pi_-}(P)$. Q.E.D.

\section{Derivation of the recurrence relations for the wave functions \eqref{lambda-}
and \eqref{lambda+}}

From the solutions of the spectral problem in terms of the
projection operators \eqref{PhifromP} we obtain a formula for
$\Psi_k=(1-\lambda)^2(\mathbb{I}-\Phi_k),\quad k=1,...,N-1$
\be
\label{psi1}
\Psi_k(\la)-\Psi_{k-1}(\la) = 2(1-\la)\left(P_k-\frac{1+\la}{1-\la}P_{k-1}\right).
\ee
If we combine the solution for $\Phi(\la)$ with that for
$[\Phi(\la)]^{-1}=\Phi(-\la)$, we simply obtain, for $l=0,...,N-1$
\be\label{psi2}
\Psi_l(\la)+\Psi_l(-\la)=4 P_l.
\ee
Substitution of \eqref{psi2} for $l=k$ and for $l=k-1$ into
\eqref{psi1} immediately yields the ``annihilation operator''
\eqref{lambda-} if we solve \eqref{psi1} for $\Psi_{k-1}$ and express
$P_{k-1}$ as $\mathbf{\Pi_-}(P_k)$. The same equations
(\ref{psi1},\ref{psi2}) yield the ``creation operator'' if we solve
\eqref{psi1} for $\Psi_{k}$ while expressing $P_{k}$ as
$\mathbf{\Pi_+}(P_{k-1})$.

\section{Derivation of equation \eqref{Pk from Xk} used in the recurrence
relations for the immersion functions \eqref{chi-} and \eqref{chi+}}

We square equation \eqref{XfromP}, bearing in mind that the
projectors ${P_0,...,P_k}$ are mutually orthogonal and we obtain
\be
X_k\c
X_k=\left[\frac{2(2k+1)}{N}-1\right]P_k
+4\left[\frac{(2k+1)}{N}-1\right]\sum\limits_{j=0}^{k-1}P_j
-\frac{(2k+1)^2}{N^2}\mathbb{I}.
\ee
This equation may be combined with $X_k$ multiplied by an appropriate
factor, as in \eqref{Pk from Xk}, to get rid of the sum of the lower
operators $P_0+...+P_{k-1}$. The solution for $P_k$ is precisely what
was found for the equation \eqref{Pk from Xk}.

\section{Derivation of the fact that the holomorphic functions $J_k$
vanish when the $\cpn$ model is defined on $S^2$ and its action functional is finite}

To prove the vanishing of the holomorphic quantities $J_k$ and
$\bar{J}_k$, it is sufficient to consider the orthogonality condition
for the operator $P_{\pm}^kf$ in the specific case where $i=k$ and
$j=k+2$\ \  \cite{WZ}
\be
\label{orth}
(P_{\pm}^kf)^{\dagger}\cdot(P_{\pm}^{k+2}f)=0.
\ee
Here $0\leq k\leq N-2$ for the operator $P_+$ or $1\leq k \leq N-1$
for the operator $P_-$. Using the notation $f_k=P_{\pm}^kf$, we get
\be
\label{orth1}
0=f_k^{\dagger}\cdot(P_{\pm}^2f_k)=f_k^{\dagger}\cdot\left(\mathbb{I}
-\frac{(P_{\pm}f_k)\otimes(P_{\pm}f_k)^{\dagger}}{(P_{\pm}f_k)^{\dagger}
\cdot(P_{\pm}f_k)}\right)\cdot\partial_{\pm}(P_{\pm}f_k)=f_k^{\dagger}
\cdot\partial_{\pm}(P_{\pm}f_k),
\ee
where the symbol $\partial_+$ represents the holomorphic derivative
$\p$  and $\partial_-$ represents the antiholomorphic derivative
$\bp$. Since $f_k^{\dagger}\cdot(P_{\pm}f_k)=0$, this implies
\be
\label{orth2}
f_k^{\dagger}\cdot\partial_{\pm}(P_{\pm}f_k)=-\partial_{\pm}f_k^{\dagger}\cdot(P_{\pm}f_k).
\ee
The right hand side of the equation (\ref{orth1}) can be written in
terms of the holomorphic function $J_k$
\be
\label{orth3}
\begin{split}
0&=-\partial_{+}f_k^{\dagger}\cdot(\mathbb{I}-P_k)\cdot\partial_{+}f_k=-(\partial_{+}P_k\cdot
P_k)_{11}=-(P_k)_{11}J_k.\\
0&=-\partial_{-}f_k^{\dagger}\cdot(\mathbb{I}-P_k)\cdot\partial_{-}f_k=-(\partial_{-}P_k\cdot
P_k)_{11}=-(P_k)_{11}\bar{J}_k.
\end{split}
\ee
Since $(P_k)_{11}\neq 0$ we get $J_k=0$. Hence $J_k$ and $\bar{J}_k$
vanish identically. The version with operator $P_+$ works for
holomorphic and mixed solutions, while the version with operator
$P_-$ works for antiholomorphic and mixed solutions. Q.E.D.


\section*{References}

\begin{footnotesize}

\end{footnotesize}

\clearpage



\end{document}